\documentclass[aps,twocolumn,nofootinbib,preprintnumbers,superscriptaddress,eqsecnum]{revtex4-1}
\usepackage{color}
\usepackage[normalem]{ulem}
\usepackage{amsmath}
\usepackage{enumerate}
\usepackage{amsfonts}
\usepackage{epsfig}

\usepackage[
      colorlinks=true,
      linkcolor=blue,
      urlcolor=blue,
      filecolor=black,
      citecolor=red,
      pdfstartview=FitV,
      pdftitle={},
        pdfauthor={},
        pdfsubject={},
        pdfkeywords={},
        pdfpagemode=None,
        bookmarksopen=true
      ]{hyperref}

\newcommand{\be}{\begin{equation}}
\newcommand{\ee}{\end{equation}}
\newcommand{\bea}{\begin{eqnarray}}
\newcommand{\eea}{\end{eqnarray}}

\newcommand{\bln}{\begin{align}}
\newcommand{\eln}{\end{align}}
\newcommand{\bst}{\begin{split}}
\newcommand{\est}{\end{split}}
\newcommand{\bi}{\begin{itemize}}
\newcommand{\ei}{\end{itemize}}
\newcommand{\ben}{\begin{enumerate}}
\newcommand{\een}{\end{enumerate}}

\def\le{\left}
\def\ri{\right}
\def\ha{{1\over 2}}
\def\lam{{\lambda}}

\def\Sig{{\Sigma}}

\def\th{{\theta}}

\def\Om{{\Omega}}
\def \th{{\theta}}

\def \lam {\lambda}
\def \om {\omega}

\newcommand{\p}{\partial}

\newcommand\ga{{\ensuremath{{\gamma}}}}

\def\lam{{\lambda}}

\def\eeq{\end{equation}}

\newcommand\sL{{\ensuremath{{\mathcal L}}}}

\newcommand\sO{{\ensuremath{{\mathcal O}}}}

\begin{document}

\title {Electric fields and quantum wormholes}

\author{Dalit Engelhardt}
\email{engelhardt@physics.ucla.edu}
\affiliation{Department of Physics and Astronomy, University of California, Los Angeles, CA 90095, USA}
\affiliation{Institute for Theoretical Physics, University of Amsterdam, Science Park 904, Postbus 94485, 1090 GL Amsterdam, The Netherlands}

\author{Ben Freivogel}
\email{benfreivogel@gmail.com}
\affiliation{Institute for Theoretical Physics, University of Amsterdam, Science Park 904, Postbus 94485, 1090 GL Amsterdam, The Netherlands}
\affiliation{GRAPPA, University of Amsterdam, Science Park 904, Postbus 94485, 1090 GL Amsterdam, The Netherlands}

\author{Nabil Iqbal}
\email{n.iqbal@uva.nl}
\affiliation{Institute for Theoretical Physics, University of Amsterdam, Science Park 904, Postbus 94485, 1090 GL Amsterdam, The Netherlands}

\begin{abstract}
Electric fields can thread a classical Einstein-Rosen bridge.  Maldacena and Susskind have recently suggested that in a theory of dynamical gravity the entanglement of ordinary perturbative quanta should be viewed as creating a quantum version of an Einstein-Rosen bridge between the particles, or a ``quantum wormhole.'' We demonstrate within low-energy effective field theory that there is a precise sense in which electric fields can also thread such quantum wormholes. We define a non-perturbative ``wormhole susceptibility'' that measures the ease of passing an electric field through any sort of wormhole.  The susceptibility of a quantum wormhole is suppressed by powers of the $U(1)$ gauge coupling relative to that for a classical wormhole but can be made numerically equal with a sufficiently large amount of entangled matter.
%We demonstrate that within low-energy effective field theory there is a qualitative equivalence between the ability to measure an electric field through a classical Einstein-Rosen bridge and through a ``quantum wormhole'': a configuration consisting of entangled matter with no geometric connection. This lends support to the ER=EPR proposal of Maldacena and Susskind and provides a precise non-holographic sense in which entanglement alone can mimic a geometric connection in a particular observational sense. We define a non-perturbative ``wormhole susceptibility'' that measures the ease of passing an electric field through any sort of wormhole. For a quantum wormhole this susceptibility is suppressed by powers of the $U(1)$ gauge coupling relative to that for a classical wormhole; however, under conditions favorable to a black hole collapse of the entangled matter in the former case, the susceptibility is in principle not impeded from increasing up to the ER value, implying that in this case this observable may see no distinction between the classical and the non-geometric quantum case.
\end{abstract}
\maketitle

%\tableofcontents

\section{Introduction}

The past decade has seen accumulating evidence of a deep connection between classical spacetime
geometry and the entanglement of quantum fields. 
%Many of the recent and ongoing studies of both bulk reconstruction and field theory entanglement structure have been motivated by the Ryu-Takayanagi
%prescription for recasting the area and reach of extremal surfaces
%in a gravitational holographic bulk as a quantitative statement about
%entanglement entropy in a dual CFT~\cite{Ryu:2006ef,Ryu:2006bv,Hubeny:2007xt}. This prescription provides a
In the AdS/CFT context, there appears to be a 
precise \textit{holographic} sense in which a classical geometry is
``emergent'' from quantum entanglement in a dual field theory (see e.g. \cite{Maldacena:2001kr,Ryu:2006ef,Ryu:2006bv,Hubeny:2007xt, PhysRevD.86.065007,Faulkner:2013ica,Lashkari:2013koa,VanRaamsdonk:2009ar,VanRaamsdonk:2010pw,Swingle:2014uza}). 

Recently, Maldacena and Susskind have made a stronger statement: that the link between entanglement and geometry exists even without any holographic changes of duality frame ~\cite{Maldacena:2013xja}. They propose that \textit{any} entangled perturbative quantum matter in the bulk of a dynamical theory of gravity, such as an entangled Einstein-Podolsky-Rosen (EPR) \cite{Einstein:1935rr} pair of electrons, is connected by a ``quantum
wormhole,'' or some sort of Planckian, highly fluctuating, version
of the classical Einstein-Rosen (ER)~\cite{Einstein:1935tc} bridge that
connects the two sides of an eternal black hole. Notably, while it
clearly resonates well with holographic ideas \cite{Susskind:2013lpa,Chernicoff:2013iga,Susskind:2013aaa,Susskind:2014rva,Susskind:2014ira,Stanford:2014jda,Susskind:2014jwa,Susskind:2014yaa,Roberts:2014isa,Susskind:2014moa,Jensen:2013ora,Sonner:2013mba,Jensen:2014lua,Jensen:2014bpa,Gharibyan:2013aha,Papadodimas:2015jra}, this ``ER = EPR'' proposal
is more general in that it makes no reference to gauge-gravity duality. The
entangled quantum fields here exist already in a theory of dynamical gravity rather than in a holographically dual field theory. 

\begin{figure}[h]
\begin{center}
\includegraphics[scale=0.27]{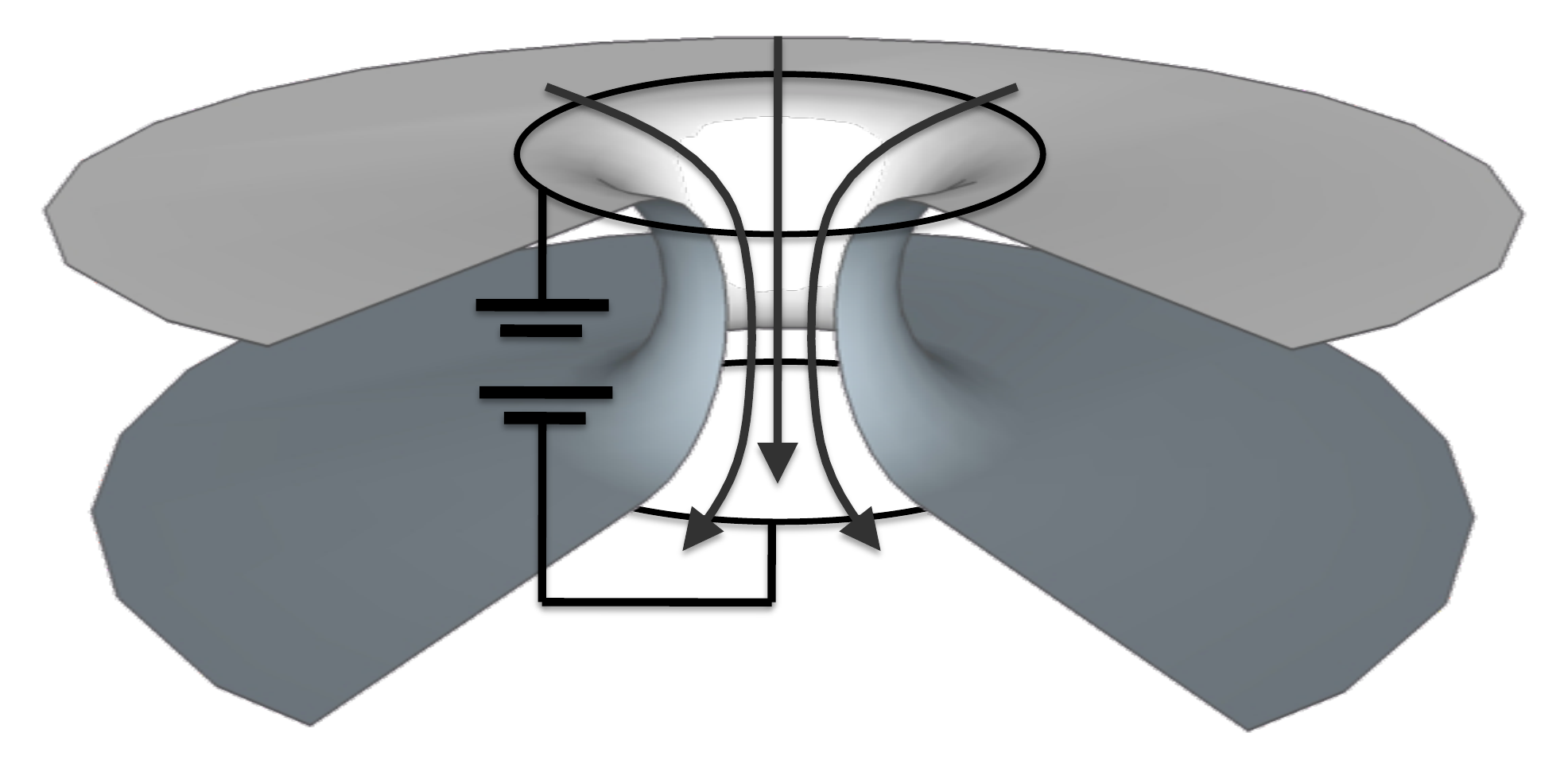}
\includegraphics[scale=0.24]{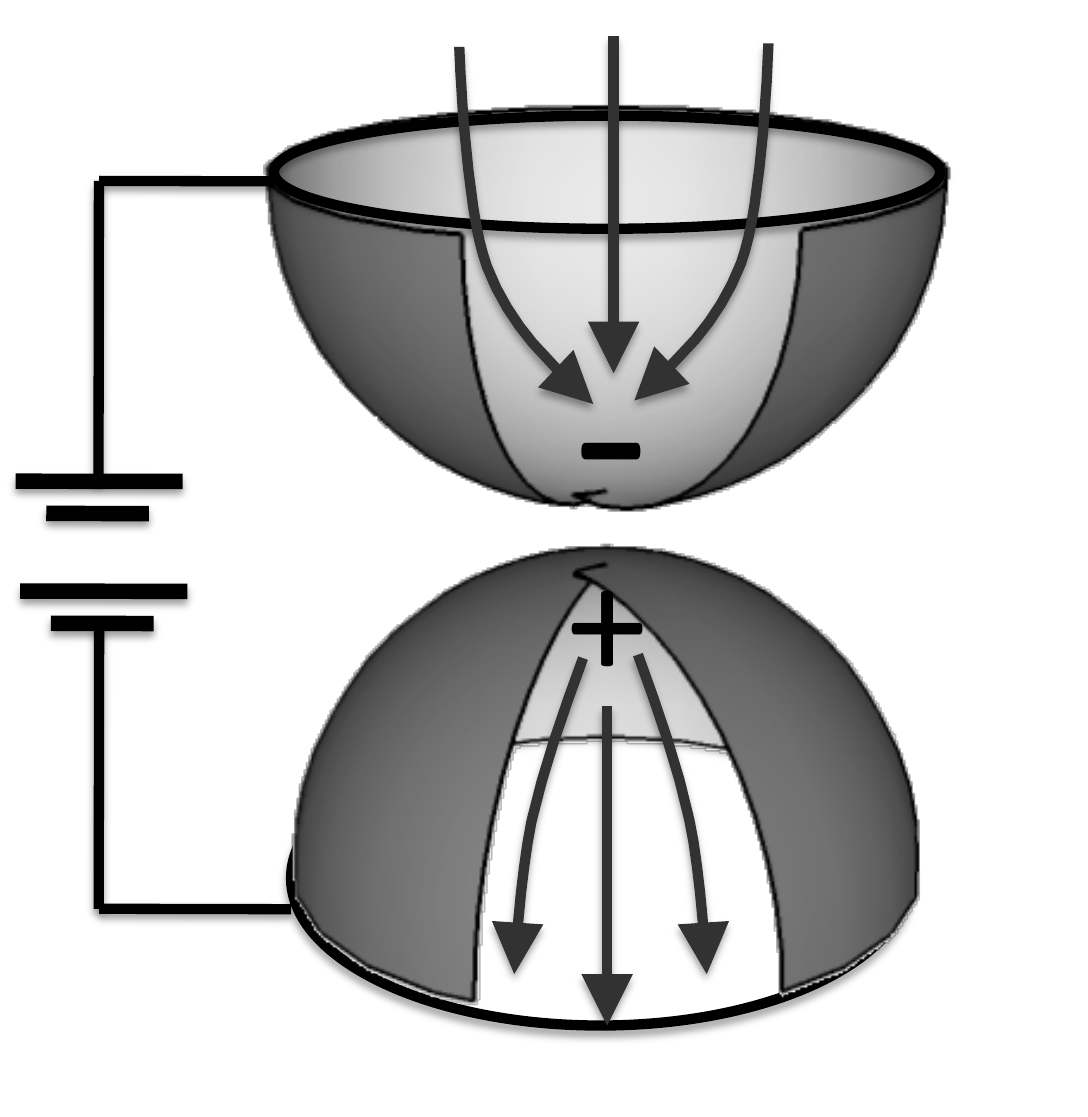}
\end{center}
\vskip -0.5cm
\caption{{\it Left:} a classical Einstein-Rosen bridge with an applied potential difference has a tunable electric field threading it. {\it Right:} A ``quantum wormhole'', -- i.e. charged perturbative matter prepared in an entangled state, with no explicit geometric connection between the two sides also has a qualitatively similar electric field threading it.}\label{fig:erbridge}
\end{figure}

It is not at all obvious that quantum wormholes so defined -- i.e. just ordinary entangled perturbative matter -- exhibit properties similar to those of classical wormholes. For example, if we have dynamical electromagnetism, then the existence of a smooth geometry in the throat of an Einstein-Rosen bridge means that there exist states with a continuously tunable\footnote{Note that here and throughout the rest of the paper, the word ``tunable'' means only that there exists a family of states with a continuously variable flux. The flux cannot actually be tuned by any local observer, as the two sides of the wormhole are causally disconnected. There does exist a quantum tunneling process in which such flux-threaded black holes can be created from the vacuum in the presence of a strong electric field \cite{Garfinkle:1990eq}.} electric flux threading the wormhole, as shown in Figure \ref{fig:erbridge}. Wheeler has described such states as ``charge without charge'' \cite{Misner:1957mt}. 

On the other hand, the two ends of a quantum wormhole may be entangled but are not connected by a smooth geometry. One might naively expect that Gauss's law would then preclude the existence of states with a continuously tunable electric flux through the wormhole. The main point of this paper is to demonstrate that this intuition is misleading: we will show that a quantum wormhole,
made up of only entangled (and charged) perturbative matter, also
permits electric fields to thread it in a manner that to distant observers with access to information about both sides, appears qualitatively the same as that for a classical ER bridge.

To quantify this, for any state $|\psi \rangle$ of either system we define a dimensionless quantity called the {\it wormhole susceptibility} $\chi_{\Delta}$,
\be
\chi_{\Delta} \equiv \langle \psi | \Phi_{\Delta}^2 | \psi \rangle \label{suscdef}
\ee
with $\Phi_{\Delta}$ the electric flux through the wormhole. This quantity clearly measures fluctuations of the flux, and we show below that through linear response it also determines the flux obtained when a potential difference is applied across the wormhole. This susceptibility is a particular measure of electric field correlations across the two sides that can be interpreted as measuring how easily an electric field can penetrate the wormhole. We note that it is a global quantity: since such an electric field can never be set up by a conventional observer on one side of the black hole, and, as we show explicitly below, measuring the wormhole susceptibility requires access to information about the flux on both sides, there is no information being transmitted across the wormhole with this electric field.

In Section~\ref{sec:ERbridge} we compute this susceptibility for a classical ER bridge and in Section~\ref{sec:QuantumWormhole} for EPR entangled matter and compare the results. In Section~\ref{sec:WilsonLines} we discuss how one might pass a Wilson line through a quantum wormhole. In Section~\ref{sec:Conclusion}
we discuss what conditions the quantum wormhole should satisfy for its throat to satisfy Gauss's law for electric fields and conclude with some implications and generalizations of these findings. 

Our results do not depend on a holographic description and rely purely on considerations from field theory and semiclassical relativity. 

\section{Classical Einstein-Rosen bridge} \label{sec:ERbridge}
We first seek a precise understanding of what it means to have a continuously tunable electric flux through a classical wormhole. We begin with the action
\be
S = \int d^4x \sqrt{-g}\le(\frac{1}{16\pi G_N}  R - \frac{1}{4 g_F^2} F^2\ri) \ .   \label{Sgrav}
\ee
where $g_F$ denotes the $U(1)$ gauge coupling. On general grounds we expect that in any theory of quantum gravity all low-energy gauge symmetries, including the $U(1)$ above, should be {\it compact} \cite{ArkaniHamed:2006dz,Banks:2010zn}. This implies that the specification of the theory requires another parameter: the minimum quantum of electric charge, $q$. Throughout this paper we will actually work on a fixed background, not allowing matter to back-react: thus we are working in the limit\footnote{If we studied finite $G_N$, allowing for the back-reaction of the electric field on the geometry, then at the non-linear level in $\mu$ we would find instead the two-sided Reissner-Nordstrom black hole. For the purposes of linear response about $\mu = 0$ this reduces to the Schwarzschild solution studied here.} $G_N \to 0$. 

\subsection{Wormhole susceptibility}
This action admits the eternal Schwarzschild black hole as a classical solution. It has two horizons which we henceforth distinguish by calling one of them ``left'' and the other ``right''. They are connected in the interior by an Einstein-Rosen bridge \cite{Einstein:1935tc}. On each side the metric is
\be
ds^2 = -\le(1 - \frac{r_h}{r}\ri) dt^2 + \frac{dr^2}{\le(1 - \frac{r_h}{r}\ri)} + r^2 d\Om_2 \qquad r > r_h, \label{schwcoord}
\ee
and at $t = 0$ the two sides join at the bifurcation sphere at $r = r_h$. The inverse temperature of the black hole is given by $\beta = 4\pi r_h$. 

\begin{figure}[h]
\begin{center}
\includegraphics[scale=0.4]{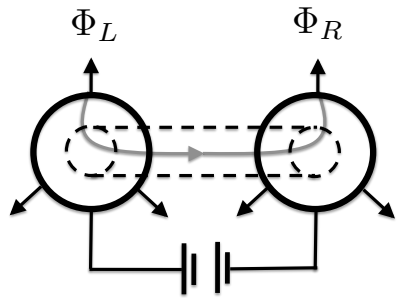}
\end{center}
\vskip -0.5cm
\caption{Electric fluxes for Einstein-Rosen bridge. Black arrows indicate sign convention chosen for fluxes. Electric field lines thread the wormhole, changing the value of $\Phi_{\Delta} \equiv (\Phi_R - \Phi_L)/2$, when a potential difference is applied across the two sides.}\label{fig:erarrows}
\end{figure}

We now surround each horizon with a spherical shell of (coordinate) radius $a > r_h$. Consider the net electric field flux through each of these spheres:
\be
\Phi_{L,R} \equiv \frac{1}{g_F^2} \int_{S^2} d\vec{A} \cdot \vec{E}_{L,R},
\ee
where the orientation for the electric field on the left and right sides is shown in Figure \ref{fig:erarrows}. 

There is an important distinction between the {\it total} flux and the {\it difference} in fluxes, defined as
\be
\Phi_{\Sig} \equiv \Phi_R + \Phi_L \qquad \Phi_{\Delta} \equiv \ha(\Phi_R - \Phi_L) \ . 
\ee
Via Gauss's law, the total flux $\Phi_{\Sig}$ simply counts the total number of charged particles inside the Einstein-Rosen bridge. It is ``difficult'' to change, in that changing it actually requires the addition of charged matter to the action \eqref{Sgrav}. Furthermore it will always be quantized in units of the fundamental electric charge $q$. 

On the other hand, $\Phi_{\Delta}$ measures instead the electric field {\it through} the wormhole. It appears that it can be continuously tuned. 

We present a short semiclassical computation to demonstrate what we mean by this. We set up a potential difference $V = 2\mu$ between the left and right spheres by imposing the boundary conditions $A_t(r_R = a) = \mu$, $A_t(r_L = a) = -\mu$. This is a capacitor with the two plates connected by an Einstein-Rosen bridge. The resulting electric field in this configuration can be computed by solving Maxwell's equations, which are very simple in terms of the conserved flux
\be
\Phi  = \frac{1}{g_F^2} \int d^2\Omega_2\;r^2 F_{rt} \qquad \p_r \Phi = 0 \ . 
\ee 
As there is no charged matter all the different fluxes are equivalent: $\Phi_R = - \Phi_L = \Phi_{\Delta}$. By symmetry we have $A_t(r_h) = 0$, and so we have
\be
\mu = A_t(r_R = a) = \int_{r_h}^a dr F_{rt} = \Phi\frac{g_F^2}{4\pi}\le(\frac{1}{a} - \frac{1}{r_h}\ri) \ . 
\ee 
We now take $a \to \infty$ for simplicity to find:
\be
\Phi_{\Delta} = \le(\frac{4\pi r_h}{g_F^2}\ri)\mu \label{fluxBH}
\ee 
As we tune the parameter $\mu$, $\Phi_{\Delta}$ changes continuously as we pump more electric flux through the wormhole. There is thus a clear qualitative difference between $\Phi_{\Delta}$ and the quantized total flux $\Phi_{\Sig}$. In this system the difference arose entirely from the fact that there is a geometric connection between the two sides. 

We now seek a quantitative measure of the strength of this connection. The prefactor relating the flux to $\mu$ in \eqref{fluxBH} is a good candidate. To understand this better, we turn now to the full quantum theory of $U(1)$ electromagnetism on the black hole background. The prefactor is actually measuring the fluctuations of the flux $\Phi_{\Delta}$ around the ground state of the system and is equivalent to the {\it wormhole susceptibility} defined in \eqref{suscdef}:
\be
\chi_{\Delta} \equiv \langle \psi | \Phi_{\Delta}^2 |\psi \rangle \label{quantdef} \ . 
\ee
To see this, recall that we are studying the Hartle-Hawking state for the Maxwell field. When decomposed into two halves (with $\mu = 0$) this state takes the thermofield double form~\cite{Israel:1976ur,Maldacena:2001kr}:
\be
|\psi \rangle \equiv \frac{1}{\sqrt{Z}} \sum_{n} |n^* \rangle_L |n\rangle_R \exp\le(-\frac{\beta}{2}E_n\ri) \ .  \label{defTf}
\ee 
Here $L$ and $R$ denote the division of the Cauchy slice at $t = 0$ into the left and right sides of the bridge, $n$ labels the exact energy eigenstates of the Maxwell field, $E_n$ denotes the energy with respect to Schwarzschild time $t$, and $|n^*\rangle$ is the CPT conjugate of $|n\rangle$ \footnote{The pairing of a state $|n\rangle$ with its CPT conjugate $|n^{\star}\rangle$ can be understood as following from path-integral constructions of the thermofield state by evolution in Euclidean time.}. 

The two-sided black hole has a non-trivial bifurcation sphere $S^2$. The electric flux through this $S^2$ is a quantum degree of freedom that can fluctuate. In the decomposition above we have two separate operators $\Phi_{L,R}$, both of which are conserved charges with discrete spectra, quantized in units of $q$: $\Phi = q \mathbb{Z}$. Each energy eigenstate can be picked to have a definite flux $\Phi_n$: $|n, \Phi_n\rangle$. Importantly, CPT preserves the energy but flips the sign of the flux.  Schematically, we have:
\be
\mbox{CPT}|n, \Phi_n\rangle = |n, -\Phi_n\rangle \ .
\ee
This means that each $L$ state in the sum \eqref{defTf} is paired with an $R$ state of opposite flux, and so the state is annihilated by $\Phi_L + \Phi_R$:
\be
(\Phi_L + \Phi_R)|\psi \rangle = 0 \label{gaussop}
\ee
This relation is Gauss's law: every field line entering the left must emerge from the right.

\begin{figure}[h]
\begin{center}
\includegraphics[scale=0.3]{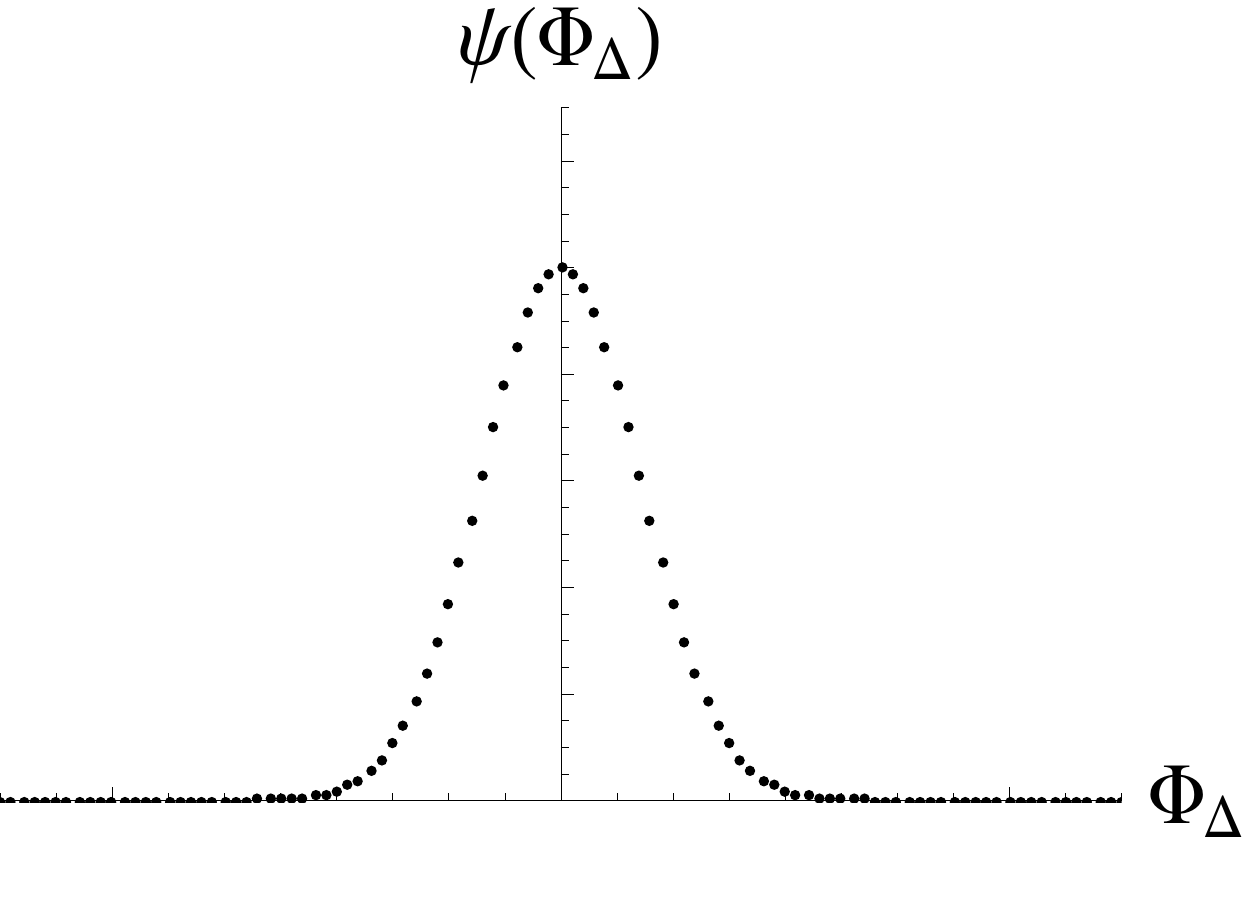}
\includegraphics[scale=0.3]{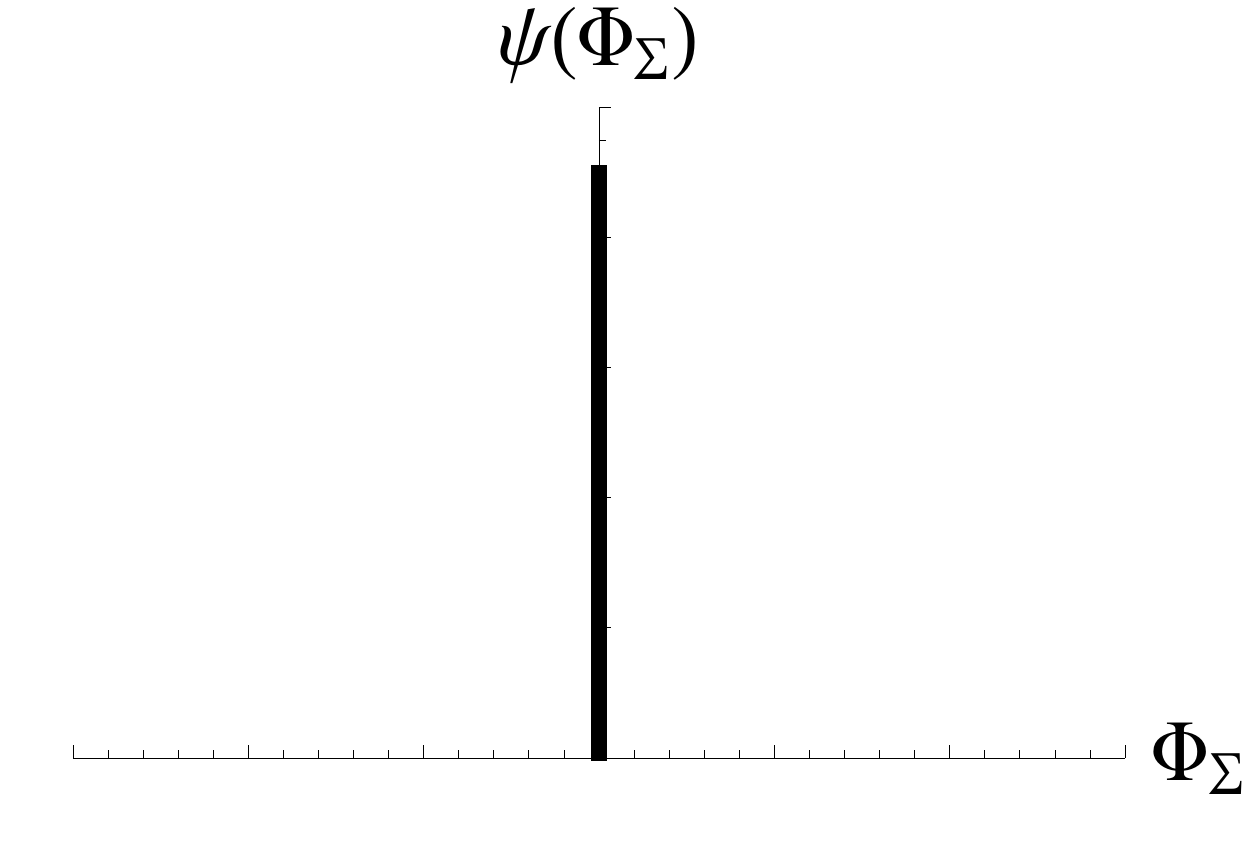}
\end{center}
\vskip -0.5cm
\caption{Wavefunction of Maxwell state as a function of discrete fluxes $\Phi_{\Delta}$ and $\Phi_{\Sig}$. The spread in $\Phi_{\Delta}$ is measured by the wormhole susceptibility $\chi_{\Delta}$. The wavefunction has no spread in $\Phi_{\Sig}$.}\label{fig:wavefuncs}
\end{figure}

On the other hand, $\Phi_{\Delta}$ does {\it not} have a definite value on this state: as $\Phi_{\Delta}$ does not annihilate $|\psi \rangle$, the wavefunction has a spread centered about zero, as shown in Figure \ref{fig:wavefuncs}. The spread of this wavefunction is measured by the wormhole susceptibility \eqref{quantdef}. The intuitive difference between $\Phi_{\Sig}$ and $\Phi_{\Delta}$ discussed above can be traced back to the fact that the wavefunction is localized in the former case and extended in the latter. 

The location of this maximum is not quantized and can be continuously tuned. For example, let us deform \eqref{defTf} with a chemical potential $\mu$:
\be
|\psi(\mu) \rangle \equiv \frac{1}{\sqrt{Z}} \sum_{n} |n^* \rangle_L |n\rangle_R \exp\le(-\frac{\beta}{2}\le(E_n - \mu \Phi_n\ri)\ri) \ . \label{defTfmu}
\ee
Similarly deformed thermofield states and the existence of flux through the Einstein-Rosen bridge have been recently studied in \cite{Andrade:2013rra,Leichenauer:2014nxa,Harlow:2014yoa}. Expanding this expression to linear order in $\mu$ we conclude that
\be
\langle \psi(\mu) |\Phi_{\Delta}|\psi(\mu)\rangle = \beta \mu\chi_{\Delta}  + \cdots \label{pert}
\ee
with $\chi_{\Delta}$ the wormhole susceptibility \eqref{quantdef} evaluated on the undeformed state \eqref{defTf}. This expectation value is the precise statement of what was computed semiclassically in \eqref{fluxBH}\footnote{More precisely: the classical relation \eqref{fluxBH} amounts to a saddle-point evaluation of a particular functional integral which evaluates expectation values of \eqref{defTf}.}: comparing these two relations we see that the wormhole susceptibility for the black hole is
\be
\chi_{\Delta}^{\mathrm{ER}} = \frac{1}{g_F^2} \ . \label{ERsusc1}
\ee

\subsection{Quantization of flux sector on black hole background}
It is instructive to provide a more explicit derivation of \eqref{ERsusc1} by computing the full wavefunction as a function of $\Phi_{\Delta}$. This requires the determination of the energy levels $E_n$ in (\ref{defTfmu}). As we are interested in the {\it total} flux, we need only determine an effective Hamiltonian describing the quantum mechanics of the flux sector. We ignore fluctuations in $A_{\th, \phi}$ and any angular dependence of the fields, integrating over the $S^2$ in \eqref{Sgrav} to obtain the reduced action:
\be
S = -\frac{2\pi}{g_F^2} \int dr dt \sqrt{-g} g^{rr}g^{tt} (F_{rt})^2 \ . 
\ee
To compute the $E_n$ we pass to a Hamiltonian formalism with respect to Schwarzschild time $t$. We first consider the Hilbert space of the right side of the thermofield double state~\eqref{defTf}. The canonical momentum conjugate to $A_r$ is the electric flux:
\be
\Phi \equiv \frac{\delta \sL}{\delta \p_t A_r} = \frac{4\pi}{g_F^2} \sqrt{-g} F^{rt} \ . 
\ee
$A_t$ does not have a conjugate momentum. The Hamiltonian is constructed in the usual way as $H \equiv \Phi \p_t A_r - \sL$ and is
\begin{align}
H & = \int_{r_h}^{\infty} dr \le(-\frac{g_F^2}{8 \pi \sqrt{-g} g^{rr} g^{tt}} \Phi^2 - (\p_r \Phi) A_t\ri) \nonumber \\ 
& + \Phi (A_t(r_h) - A_t(\infty))
\end{align}
where we have integrated by parts. The equation of motion for $A_t$ is Gauss's law, setting the flux to a constant: $\p_r \Phi = 0$. 

There are two boundary terms of different character. The value of $A_t(\infty) \equiv \mu$ at infinity is set by boundary conditions. If $\mu$ is nonzero, then the Hamiltonian is deformed to have a chemical potential for the flux as in \eqref{defTfmu}. Recall, however, that the susceptibility is defined in the undeformed state, as in \eqref{suscdef}. For the remainder of this section we therefore set $\mu$ to zero. On the other hand, $A_t(r_h)$ is a dynamical degree of freedom. We should thus combine this Hamiltonian with the corresponding one for the left side of the thermofield state; demanding that the variation of the horizon boundary term with respect to $A_t(r_h)$ vanishes then requires that $\Phi_L = -\Phi_R$, as expected from \eqref{gaussop}.

As the flux is constant we may now perform the integral over $r$ to obtain the very simple Hamiltonian
\be
H = G \Phi^2 \qquad G \equiv - \int_{r_h}^{\infty} dr \frac{g_F^2}{8 \pi \sqrt{-g} g^{rr} g^{tt}} = \frac{g_F^2}{8\pi r_h} \label{ham}
\ee
This Hamiltonian describes the energy cost of fluctuations of the electric field through the horizon of the black hole. 

The flux operator in the reduced Hilbert space\footnote{Where, as above, we neglect fluctuations along the angular directions.} of the flux sector acts as 
\be
\Phi|m\rangle = qm | m \rangle \qquad m \in \mathbb{Z}
\ee
where $m\in\mathbb{Z}$ denotes the number of units of flux carried by each state $| m \rangle$. The Hamiltonian \eqref{ham} is diagonal in this flux basis, with the energy of a state with $m$ units of flux given by
\be
E_m = \frac{g_F^2}{8\pi r_h} (qm)^2 \qquad m \in \mathbb{Z}
\ee
Though we do not actually need it, for completeness we note that the operator that changes the value of the flux through the $S^2$ is a spacelike Wilson line that pierces it carrying charge $q$:
\be
W = \exp\le(i q \int dr A_r\ri) \ . 
\ee 
Indeed from the fundamental commutation relation $[A_r(r), \Phi(r')] = i \delta(r-r')$ we find the commutator
\be
[W, \Phi] = -q W,
\ee
meaning that a Wilson line that pierces the $S^2$ once increases the flux by $q$. In our case any Wilson line that pierces the left sphere must continue to pierce the right: thus if it increases the left flux it will decrease the right flux, and we are restricted to the gauge-invariant subspace that is annihilated by $\Phi_L + \Phi_R$. 

Thus we see that the thermofield state \eqref{defTf} in the flux sector takes the simple form:
\be
|\psi \rangle = \frac{1}{\sqrt{Z}} \sum_{m \in \mathbb{Z}} |-m \rangle |m \rangle \exp\le(-\frac{g_{F}^2}{4} (qm)^2\ri), \label{fluxstates}
\ee
where we have used $\beta = 4\pi r_h$. Since $\Phi_{\Delta} = \Phi_R$ on this state, the probability of finding any flux $\Phi_{\Delta}$ through the black hole is simply
\be
P(\Phi_{\Delta}) = \frac{1}{Z} \exp\le(-\frac{g_F^2}{2}\Phi_{\Delta}^2\ri) \ . \label{prob}
\ee
This is the (square of the) wavefunction shown in Figure \ref{fig:wavefuncs}: even though $\Phi_{\Delta}$ has a discrete spectrum, the wavefunction is extended in $\Phi_{\Delta}$. In the semiclassical limit $(g_F q) \to 0$ the discreteness of $\Phi_{\Delta}$ can be ignored and the spread $\chi_{\Delta} = \langle \Phi_{\Delta}^2 \rangle$ is again $\chi_{\Delta}^{\mathrm{ER}} = g_F^{-2}$,
in agreement with the result found from the classical analysis \eqref{ERsusc1}. 

The probability distribution exhibited in \eqref{prob} may be surprising: we are asserting that an observer hovering outside an uncharged eternal black hole nevertheless finds a nonzero probability of measuring an electric flux through the horizon. However, due to \eqref{gaussop} the flux measured by the right observer will always be precisely anti-correlated with that measured by the left observer. These observers are measuring fluctuations of the field {\it through} the wormhole, not fluctuations of the number of charges inside. Through \eqref{pert} we see that it is actually the presence of these fluctuations that makes it possible to tune the electric field through the wormhole. In the above analysis, we have computed the fluctuations in the Hartle-Hawking state; more generally, any nonsingular state of the gauge fields in the ER background will have correlated fluctuations in the flux, arising from the correlated electric fields near the horizon.

\section{Quantum wormhole}\label{sec:QuantumWormhole}

We now consider the case of charged matter in an entangled state but with no geometrical, and hence no gravitational, connection. 
We will show below that when we apply a potential difference, an appropriate pattern of entanglement between the boxes is sufficient to generate a non-vanishing electric field even though the two boxes are completely noninteracting.

%The electric flux $\Phi_{\Delta}$ is thus seen to arise purely from the entanglement of the fields when a potential difference between the boxes is applied.

\begin{figure}[h]
\begin{center}
\includegraphics[scale=0.4]{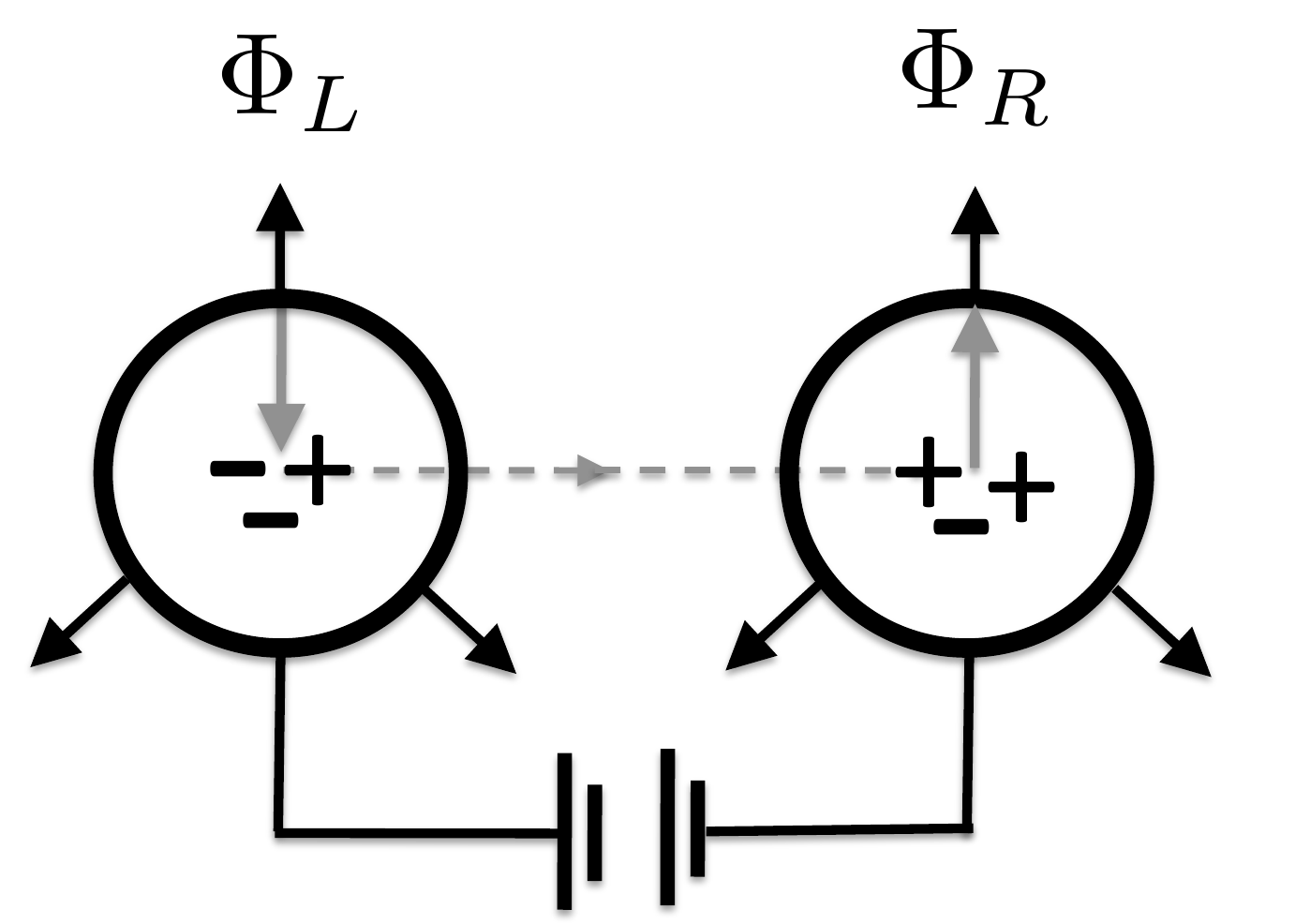}
\end{center}
\vskip -0.5cm
\caption{Setup for quantum wormhole. The two boxes are geometrically disconnected but contain a scalar field in an entangled state. Correlated charge fluctuations effectively allow electric field lines to travel from one box to the other when a potential difference is applied.}\label{fig:eprarrows}
\end{figure}

The configuration that we study here is that of a complex scalar field $\phi$ charged under a $U(1)$ symmetry (with elementary charge $q$), confined to two disconnected spherical boxes of radius $a$, as shown in Figure \ref{fig:eprarrows}. The confinement to $r<a$ is implemented by imposing Dirichlet boundary conditions for the fields. These boundary conditions still allow the radial electric field to be nonzero at the boundary, so our main observable, the electric flux, is not constrained by the boundary conditions.

The action in each box is
\be
S = \int d^4x \sqrt{-g}\le(-|D\phi|^2 - m^2 \phi^{\dagger}\phi - \frac{1}{4g_F^2} F^2\ri) \ . 
\ee
where $D_{\mu}\phi=\partial_{\mu}\phi-iqA_{\mu}\phi$. This is now an interacting theory where the perturbative expansion is controlled by $(g_F q)^2$. 

Gauss's law relates the electric flux to the total global charge $Q$. Thus we have the following operator equation on physical states:
\be
\Phi = \frac{1}{g_F^2}\int d^3x \le(\nabla \cdot E\ri ) = Q \label{gaussQ}.
\ee
At first glance this situation is rather different from the classical black hole case. $\Phi_L$ and $\Phi_R$ simply measure the number of particles in the left and right boxes respectively. There appears to be no difference between $\Phi_{\Sig}$ and $\Phi_{\Delta}$ and thus no way to thread an electric field through the boxes. This intuition is true in the vacuum of the field theory, which is annihilated by both $\Phi_L$ and $\Phi_R$. As it turns out, it is wrong in an entangled state. 

Let us now perform the same experiment as for the black hole: we will set up a potential difference of $2\mu$ between the two spheres by studying the analog of the deformed thermofield state \eqref{defTfmu}. We will work at weak coupling: the only effect of the nonzero coupling is to relate the flux to the global charge as in \eqref{gaussQ}. We are thus actually studying charge fluctuations of the scalar field. These charge fluctuations source electric fields which cost energy, but this energetic penalty can be neglected at lowest order in the coupling\footnote{It is interesting to note that in the black hole case the key difference is that the energy cost associated to the gauge fields -- which we neglect in this case -- is the {\it leading} effect.}. 

The full state for the combined Maxwell-scalar system is formally the same as \eqref{defTfmu}. We schematically label the scalar field states by their energy and global charge as $|n, Q_n\rangle$. Due to the constraint \eqref{gaussQ}, the scalar field sector of the thermofield state can be written:
\be
|\psi(\mu) \rangle \equiv \frac{1}{\sqrt{Z}} \sum_{n} |n, -Q_n \rangle_L |n, Q_n\rangle_R \exp\le(-\frac{\beta}{2}\le(E_n - \mu Q_n\ri)\ri) \label{tfphi}
\ee
This state corresponds to having a constant value of $A_t = \mu$ in the right sphere and $A_t = -\mu$ in the left sphere. Note that we have
\be
(\Phi_L + \Phi_R)|\psi(\mu)\rangle = (Q_L + Q_R)|\psi(\mu)\rangle = 0 \label{gaussphi} \ . 
\ee
We now seek to compute $\langle \Phi_\Delta \rangle_{\mu} = \langle Q_{\Delta} \rangle_{\mu} = \langle Q_R \rangle_{\mu}$ where the second equality follows from \eqref{gaussphi}. However to compute $Q_R(\mu)$ we can trace out the left side. Tracing out one side of the thermofield state results in a thermal density matrix for the remaining side: thus we are simply performing a normal statistical mechanical computation of the charge at finite temperature and chemical potential. Details of this standard computation are in Appendix \ref{sec:chargedsf}, and the result is: 
\be
\langle \Phi_{\Delta} \rangle_{\mu} = q^2 \sum_n \le(\frac{1}{1-\cosh(\beta \om_n)}\ri)(\beta \mu) + \sO(\mu^2), \label{fieldphi}
\ee 
where the $\om_n$ are the single-particle energy levels. The sum can be done numerically. 

\begin{figure}[h]
\begin{center}
\includegraphics[scale=0.6]{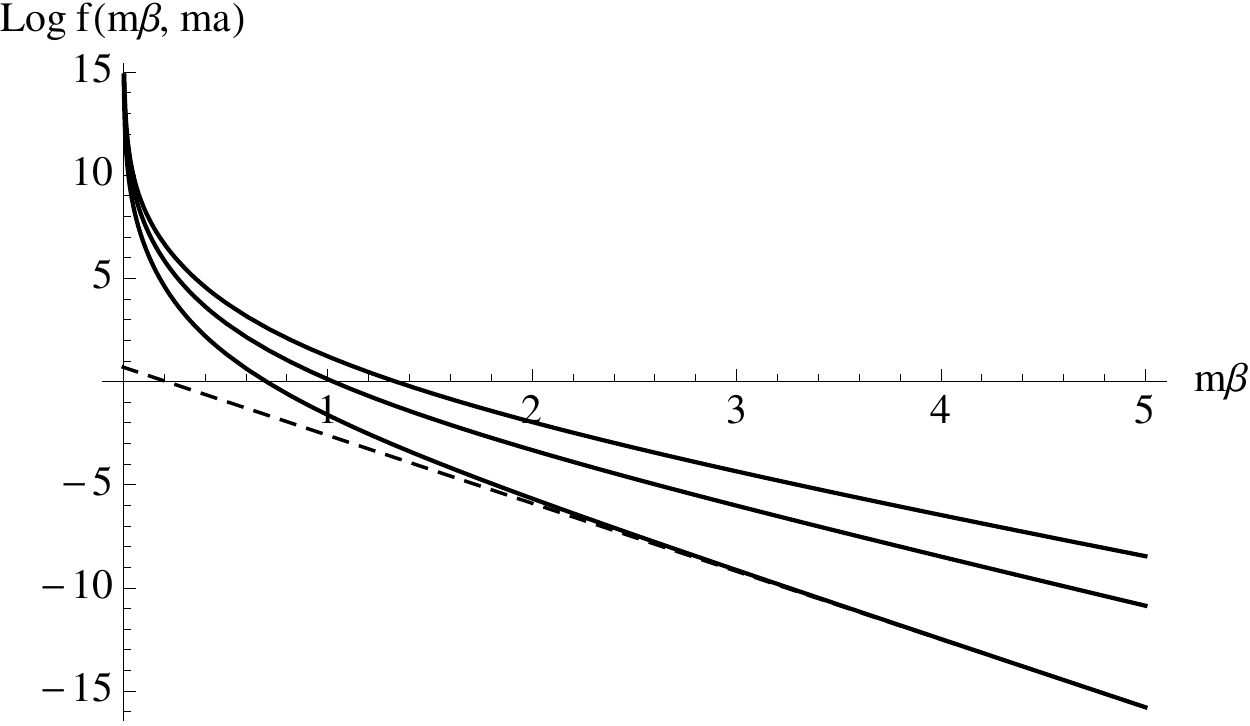}
\end{center}
\vskip -0.5cm
\caption{Numerical evaluation of logarithm of dimensionless function $f(m\beta, ma)$ appearing in wormhole susceptibility for complex scalar field. From bottom moving upwards, three curves correspond to $ma = 1, 1.5, 2$. Dashed line shows asymptotic behavior for $ma = 1$ of $\exp\le(-\om_0\beta\ri)$.}
\label{fig:susc}
\end{figure}

We conclude that the wormhole susceptibility for this state is:
\be
\chi_{\Delta}^{\mathrm{EPR}} = q^2 f\le(m\beta, ma\ri) \label{quantchi}
\ee
with $f$ a calculable dimensionless function that is $O(1)$ in the couplings and is displayed for illustrative purposes in Figure \ref{fig:susc}. Crudely speaking it measures the number of accessible charged states. If we decrease the entanglement by lowering the temperature, the susceptibility vanishes exponentially as $f \sim \exp(-\om_0\beta)$, with $\om_0$ the lowest single-particle energy level. Its precise form -- beyond the fact that it is nonzero in the entangled state -- is not important for our purposes.

We see then that as a result of the potential difference that we have set up between the two entangled spheres, we are able to measure flux fluctuations across the wormhole that are both continuously tunable and fully correlated with each other: this is observationally indistinct from measuring an electric field through the wormhole. This field exists not because of geometry, but rather because the electric field entering one sphere attempts to create a negative charge. Due to the entanglement this results in the creation of a positive charge in the other sphere, so that the resulting field on that side is the same electric field as the one entering the first sphere. This appears rather different from the mechanism at play for a geometric wormhole, but the key fact is that the wavefunction in the flux basis takes qualitatively the same form (i.e. that shown in Figure \ref{fig:wavefuncs}) for both classical and quantum wormholes, meaning that the universal response to electric fields is the same for both systems. 

Quantitatively, however, there is an important difference. If the function $f$ that measures the number of charged states is of $O(1)$, then the wormhole susceptibility for the quantum wormhole \eqref{quantchi} is smaller than that for the classical wormhole \eqref{ERsusc1} by a factor of $(g_F q)^2$. It is {\it much} harder -- i.e. suppressed by factors of $\hbar$ -- to push an electric field through a quantum wormhole. Alternatively, we can view \eqref{ERsusc1} as defining the value of the $U(1)$ gauge coupling in the wormhole region. In the quantum wormhole we have succeeded in creating a putative region through which a $U(1)$ gauge field can propagate, but its coupling there (as measured by \eqref{quantchi}) is large, and becomes larger as the entanglement is decreased. Notwithstanding these large ``quantum fluctuations'', the quantum wormhole does nevertheless satisfy topological constraints such as Gauss's law.

Despite this suppression, there is no obstruction in principle to making $f$ sufficiently large so that the susceptibilities can be made the same. Increasing the temperature or the size of the box will increase the number of charged states and thus increase $f$, as can be seen explicitly in Figure \ref{fig:susc}. Thus even the numerical value of the EPR wormhole susceptibility can be made equal or even greater than that of the ER bridge, although we will require a large number of charged particles to do it in a weakly coupled regime.

%As shown in~\ref{fig:susc}, $f$ grows with decreasing $m\beta$, so  it may bec Although it may be challenging to describe this mechanism dynamically and this critical temperature may be quite high, we see that quantum wormhole susceptibility can in general be made even larger than the classical ER one: the interpretation of this is that in a collapse scenario there should in principle be enough susceptibility ``phase space'' to form a classical ER bridge. This provides an intuitive manifestation of how a geometric connection emerges from a purely quantum entangled one.

\section{Wilson lines through the horizon} \label{sec:WilsonLines}
% As argued above, gauge field observations outside a classical and quantum wormhole appear qualitatively the same. This suggests that additional probes are needed for more refined information about the wormhole. 

It was argued above that electric flux measurements behave qualitatively the same for a classical and for a quantum wormhole. It is interesting to consider other probes involving the gauge field. For example, the classical eternal black hole also allows a Wilson line to be threaded through it. Such Wilson lines have recently been studied in a toy model of holography in \cite{Harlow:2014yoa}. For the black hole, consider extending a Wilson line from the north pole of the left sphere at $r = a$ through the horizon to the north pole of the right sphere:
\be
W_{ER} = \bigg\langle \exp\le(iq \int_{L}^R A \ri) \bigg\rangle \approx 1 \label{WilsonBH}
\ee

Since we assume that the gauge field is weakly coupled throughout the geometry, to leading order we can simply set it to $0$, leading to the approximate equality above. 

On the other hand, in the entangled spheres case where there is no geometric connection it is not clear how a Wilson line may extend from one box to the other. However, an analogous object with the same quantum numbers as \eqref{WilsonBH} is
\be
W_{EPR} = \bigg\langle\exp\le(iq \int_L^0 A_L\ri)\phi_L(0)\phi_R^{\dagger}(0)\exp\le(iq \int_0^R A_R\ri) \bigg\rangle, \label{wepr}
\ee
where each Wilson line extends now from the skin of the sphere to the center of the sphere at $r = 0$, where it ends on a charged scalar field insertion. 

While the gauge field may be set to zero as in the black hole case above, we must furthermore account for the mixed correlator of the scalar field, which is nonzero only due to entanglement. Details of the computation and a plot of the results can be found in Appendix \ref{sec:chargedsf}. The leading large $\beta$ behavior is
\be
W_{EPR} \sim \exp\le(-\frac{\om_0\beta}{2}\ri) \ .
\ee
As the temperature is decreased the expectation value of the Wilson line vanishes, consistent with the idea put forth above that the gauge field living in the wormhole is subject to strong quantum fluctuations which become stronger, washing out the Wilson line, as the entanglement is decreased.

\section{Conclusion} \label{sec:Conclusion}
A classical geometry allows an electric field to be passed through it. In this paper we have demonstrated that we can mimic this aspect of a geometric connection using entangled charged matter alone. We also introduced the wormhole susceptibility, a quantitative measure of the strength of this connection. For quantum wormholes this susceptibility is suppressed relative to that for classical wormholes by factors of the dynamical $U(1)$ gauge coupling, i.e. by powers of $\hbar$, but, as we argued in Section~\ref{sec:QuantumWormhole}, there is in principle no impediment for the susceptibilities to be of the same order. The susceptibility is defined for any state, but for the thermofield state it directly measures the electric field produced when a potential difference is applied across the wormhole. 

We stress that we are not claiming that there is a smooth geometry in the quantum wormhole; however, there is a crude sense in this setup in which a geometry emerges from the presence of entanglement. We showed that by adjusting the parameters, two boxes of weakly coupled, entangled charged particles can mimic an Einstein-Rosen bridge in their response to electric potentials. The structure we have found in this highly excited state is similar to that of the vacuum of a two-site $U(1)$ lattice gauge theory, where entanglement between the sites allows the electric flux between them to fluctuate. This is enough to allow for a nonzero susceptibility, and is somewhat reminiscent of ideas of dimensional deconstruction \cite{ArkaniHamed:2001ca}. 

The equivalence is only at the coarse level of producing the same electric flux at the boundary of the boxes; more detailed observations inside the boxes would quickly reveal that charged matter rather than black holes are present. 
However, the presence of a nonzero wormhole susceptibility already at weak coupling, along with the fact that the two susceptibilities become similar as the coupling is increased, is compatible with the idea that as we go from weak to strong coupling entangled matter becomes an ER bridge.  
 
It is worth noting that the quantum wormholes considered still satisfy Gauss's law \eqref{gaussphi} in that every field line entering one side must exit from the other. This is due to the correlated charge structure of the states considered:
\be
|\psi \rangle \sim \sum_Q |-Q \rangle |Q \rangle
\ee
If we coherently increase the charge of the left sector relative to the right, then this would correspond to having a definite number of charged particles inside the wormhole. 

Alternatively, we could consider a more generic state, involving instead an incoherent sum over all charges. 
%It has been argued on general grounds that in a generic state any two-sided correlations are suppressed by powers of the inverse density of states $e^{-S}$ \cite{Marolf:2013dba,Balasubramanian:2014gla}. The spreads $\langle \Phi_{\Sigma}^2 \rangle$ and $\langle \Phi_{\Delta}^2 \rangle$ differ only by  two-sided terms of the form $\langle \Phi_L \Phi_R \rangle$, which are vanishingly small at large entropy. There is then no distinction between these two susceptibilities, $\langle \Phi_{\Sigma}^2 \rangle \sim \langle \Phi_{\Delta}^2 \rangle$, and a
This type of generic quantum wormhole does not satisfy any analog of Gauss's law. At first glance, this appears non-geometric, in agreement with the intuition that a generic state should not have a simple geometric interpretation \cite{Marolf:2013dba,Balasubramanian:2014gla}. On the other hand, we could also simply state that we have filled the wormhole with matter that is not in a charge eigenstate, i.e. a superconducting fluid. Thus some ``non-geometric'' features nevertheless have an interpretation in terms of effective field theory, and a two-sided analog of the holographic superconductor \cite{Gubser:2008px,Hartnoll:2008vx,Hartnoll:2008kx} might capture universal aspects of the gauge field response of such a state.

We note also that the susceptibility is constructed from conserved charges, and so it commutes with the Hamiltonian. Thus the time-evolved versions of the thermofield state (which have been the subject of much recent study as examples of more ``generic'' states \cite{Roberts:2014isa,Maldacena:2013xja,Hartman:2013qma,Papadodimas:2015xma,Shenker:2013pqa,Papadodimas:2015jra}) do not scramble charges: they all have the same wormhole susceptibility as the original thermofield state and precisely satisfy Gauss's law. We also find that the wormhole susceptibility must be conserved if two disconnected clouds of entangled matter are collapsed to form two black holes, which presumably then must have an Einstein-Rosen bridge between them. This provides a crude realization of the collapse experiment proposed in \cite{Maldacena:2013xja}.

It is of obvious interest to generalize our considerations to gravitational fields. In that case the gravitational susceptibility corresponding to \eqref{ERsusc1} directly measures Newton's constant in the wormhole throat. Note also that the form of the ``ER = EPR'' correspondence studied here requires the existence of perturbative matter charged under every low-energy gauge field: e.g. to form a quantum wormhole that admits magnetic fields, we would require entangled magnetic monopoles. If the charge spectrum were not complete, one could certainly tell the difference between an ER bridge and an EPR one. Precisely such a completeness of the charge spectrum in consistent theories of quantum gravity has been conjectured on (somewhat) independent grounds \cite{ArkaniHamed:2006dz,Banks:2010zn, Polchinski:2003bq}. 

%Finally, it is remarkable that the two computations performed here, which \textit{a priori} share no features with each other besides the entanglement structure of their state, generate qualitatively similar results for the observable considered here. An interesting direction is to increase the ``resolution" of observables comparing these two setups in order to develop a more refined notion of the general character of a wormhole. This refinement could in principle lead to a greater distinction between wormholes arising from purely entanglement versus those that are also geometrically connected. Nonetheless, if we are to adopt the view that geometry is truly emergent from entanglement, or, alternatively and perhaps equivalently, that the two are in fact different limits of the same more general concept in quantum gravity, then a complete description of the latter would be expected to erase any such differences.

Finally, we find it intriguing that the two computations performed here result in qualitatively similar answers, but arising from different sources and at different orders in bulk couplings. One might be tempted to speculate that in a formulation of bulk quantum gravity that is truly non-perturbative these two very different computations could be understood as accessing a more general concept that reduces in different limits to either perturbative entanglement or classical geometry. It remains to be seen what this more general concept might be.

\begin{acknowledgments}
It is our pleasure to acknowledge helpful discussions with J. Camps, N. Engelhardt, D. Harlow, D. Hofman, P. Kraus, P. de Lange, L. Kabir, M. Lippert, A. Puhm, J. Santos, G. Sarosi, J. Sully, L. Susskind, D. Tong, L. Thorlacius, and B. Way. This work was supported by the NSF grant DGE-0707424, the University of Amsterdam, the Netherlands Organisation for Scientific Research (NWO), and the D-ITP consortium, a program of the NWO that is funded by the Dutch Ministry of Education, Culture and Science (OCW). N.I. would like to acknowledge the hospitality of DAMTP at the University of Cambridge while this work was in progress.
\end{acknowledgments}

\begin{appendix}
\section{Charged scalar field computations} \label{sec:chargedsf}
\begin{figure}[h]
\begin{center}
\includegraphics[scale=0.5]{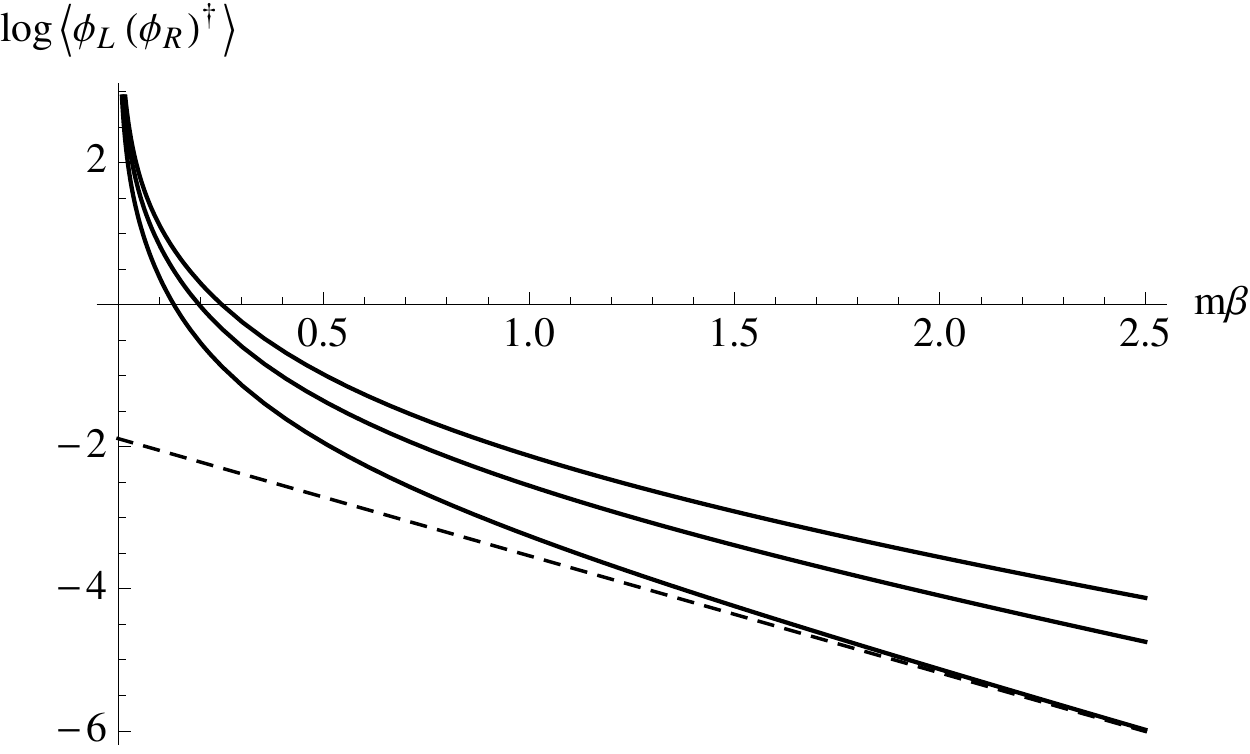}
\end{center}
\vskip -0.5cm
\caption{Numerical evaluation of logarithm of $\langle \phi_L(0) \phi_R^{\dagger}(0)\rangle$, which contributes the interesting dependence of the Wilson line \eqref{wepr}. From bottom moving upwards, curves correspond to $ma = 1, 1.5, 2$. Dashed line corresponds to asymptotic behavior for $ma = 1$ of $\exp\le(-\frac{\om_0}{2\beta}\ri)$ with $\om_0$ the lowest single-particle energy level.}
\label{fig:corr}
\end{figure}

Here we present some details of the charged scalar field computations presented in the main text. Similar results would be obtained for essentially any system in any geometry, but for concreteness we present the precise formulas for the charged scalar field in a spherical box. The relevant part of the action is
\be
S_{\phi} = -\int d^4x \le(|D\phi|^2 + m^2|\phi|^2\ri)
\ee
The scalar field is confined to a spherical box of radius $a$ with Dirichlet boundary conditions $\phi(r = a) = 0$. We first compute the single-particle energy levels. 

Expanding the field in spherical harmonics as $\phi = \sum_{lmp}\phi_{lp}(r)e^{-i\om t}\ Y_{lm}(\th,\phi)$ we find the mode equation for $\phi_{lp}(r)$ to be
\be 
\frac{1}{r^2}\p_r\le(r^2 \p_r \phi_{lp}(r)\ri) - \frac{l(l+1)}{r^2} \phi_{lp}(r) = (m^2 - \om^2) \phi_{lp}(r), \label{radialeq}
\ee
Here $p$ is a radial quantum number and $l$ is angular momentum as usual. The normalizable solutions to the radial wave equation are spherical Bessel functions of order $l$: 
\be
\phi_{lp}(r) = c_{lp} j_s(l, \lam_{lp} r) \qquad c_{lp} =  \frac{2}{\sqrt{a^3 \pi}}\le(J_{l + \frac{3}{2}}(\lam_{lp} a)\ri)^{-1} \label{modefcn}
\ee 
The normalization $c_{lp}$ has been picked such that
\be
\sum_{p} \phi_{lp}(r) \phi_{lp}(r') = \frac{\delta(r - r')}{r^2} \label{phinorm}
\ee
In $c_{lp}$, $J_{\nu}(x)$ is an ordinary Bessel function of the first kind. Imposing the Dirichlet boundary condition fixes $\lam_{p} = \frac{x_{lp}}{a}$, where $x_{lp}$ is the $p$-th zero of the $l$-th spherical Bessel function. This determines the energy levels to be
\be
\om_{lp} = \sqrt{m^2 + \le(\frac{x_{lp}}{a}\ri)^2},
\ee
We are now interested in computing the charge susceptibility at finite temperature $T$ and chemical potential $\mu$. From elementary statistical mechanics we have the usual expression for the charge 
\be
\langle Q \rangle = q \sum_{lp} (2l+1)\le(\frac{1}{1 - e^{\beta(\om_{lp} + q \mu)}} - \frac{1}{1 - e^{\beta(\om_{lp} - q\mu)}}\ri), 
\ee
where we have included the degeneracy factor $(2l+1)$. Linearizing this in $\mu$ we obtain \eqref{fieldphi}, where it is understood that the sum over single-particle states there includes a sum over angular momentum eigenstates: $\sum_n \to \sum_{lp}(2l+1)$. 

Next we compute the correlation function $\langle \phi_L^{\dagger}(0)\phi_R(0) \rangle$ across the two sides of the thermofield state \eqref{tfphi} (with $\mu \to 0$). The fastest way to compute this is to note that the two sides of the thermofield state can be understood as being connected by Euclidean time evolution through $\frac{\beta}{2}$. Thus the mixed correlator can be calculated by computing the usual Euclidean correlator between two points separated by $\frac{\beta}{2}$ in Euclidean time (see e.g. \cite{Maldacena:2001kr}). If the single-particle energy levels are given by $\om_{pl}$, then the Euclidean correlator between two general points is
\begin{align}
 G(\tau,r,\th,\phi;\;& \tau',r',\th',\phi') = \sum_{lmp}\frac{1}{2\om_{lp}} \frac{\cosh\le(\om_{lp}\le(\tau - \tau' - \frac{\beta}{2}\ri)\ri)}{\sinh\le(\frac{\beta \om_{lp}}{2}\ri)} \times \nonumber \\
& \phi_{pl}(r)\phi_{pl}(r')Y_{lm}(\th,\phi)Y_{lm}^*(\th',\phi'),
\end{align} 
where in this expression the normalization of the mode functions \eqref{phinorm} is important. 

For our application to the Wilson line in \eqref{wepr} we care about the specific case $\tau - \tau' = \frac{\beta}{2}$ and $r = r' = 0$. The spherical Bessel functions with nonzero angular momentum $l \neq 0$ all vanish at the origin $r = 0$. Thus the sum is only over the $l = 0$ modes. The result of performing this sum numerically is shown in Figure \ref{fig:corr}, but it is easy to see that at small temperatures the answer will be dominated by the lowest energy level and is:
\be
\langle \phi_L(0)^{\dagger} \phi_R(0) \rangle \sim \exp\le(-\frac{\om_0\beta}{2}\ri) \ . 
\ee
\end{appendix}

\bibliographystyle{utphys}
\bibliography{all}

\end{document}